\documentclass[aps,prb,showpacs,twoside,twocolumn,10pt]{revtex4-1}
\usepackage[colorlinks=true, citecolor=red, urlcolor=blue ]{hyperref}
\usepackage{times,epsfig,amssymb,amsfonts,amsmath,bm,subfigure,mathtools,amsthm,braket,soul,enumitem,xcolor,physics,graphics,graphicx}
\UseRawInputEncoding
\usepackage[normalem]{ulem}
\usepackage{comment}
\usepackage{epstopdf}

\begin{document}

\title{DiffCrysGen: A Score-Based Diffusion Model for Design of Diverse Inorganic Crystalline Materials}

\author{Sourav Mal$^{1,3}$, Subhankar Mishra$^{2,3}$, Prasenjit Sen$^{1,3}$}
\affiliation{$^1$Harish-Chandra Research Institute, Chhatnag Road, Jhunsi, Prayagraj 211019, India}
\affiliation{$^2$National Institute of Science Education and Research (NISER), Jatni 752050, Odisha, India}
\affiliation{$^3$Homi
Bhabha National Institute, Training School Complex, Anushakti Nagar, Mumbai 400094, India}

\begin{abstract}

Crystal structure generation is a foundational challenge in materials discovery, particularly in designing functional inorganic crystalline materials with desired properties. Most existing diffusion-based generative models for crystals rely on complex, hand-crafted priors and modular architectures to separately model atom types, atomic positions, and lattice parameters. These methods often require customized diffusion processes and conditional denoising, which can introduce additional model complexities and inconsistencies. Here we introduce DiffCrysGen, a fully data-driven, score-based diffusion model that jointly learns the distribution of all structural components in crystalline materials. With crystal structure representation as unified 2D matrices, DiffCrysGen bypasses the need for task-specific priors or decoupled modules, enabling end-to-end generation of atom types, fractional coordinates, and lattice parameters within a single framework. Our model learns crystallographic symmetry and chemical validity directly from large-scale datasets, allowing it to scale to complex materials discovery tasks. As a demonstration, we applied DiffCrysGen to the design of rare-earth-free magnetic materials with high saturation magnetization, showing its effectiveness in generating stable, diverse, and property-aligned candidates for sustainable magnet applications.

\end{abstract}

\maketitle

\section{Introduction} 

Designing novel materials with desired properties is one of the most challenging problems in materials science. This is because the material design space is astronomically large, and the specific knowledge of which periodic atomic arrangements will yield the desired properties is not known a priori. 
Traditionally, materials discovery has relied heavily on experimentation and human intuition, often involving modifications to known compounds, such as elemental substitutions or dopants introduced at specific crystallographic sites. Properties of these candidate materials are then evaluated using density functional theory (DFT) calculations in high-throughput workflows. However, the primary bottleneck in most material design pipelines is the high computational expense of DFT calculations.

With the advent of high-quality materials databases~\cite{MP,OQMD,JARVIS,aflow,Alex-1,Alex-2,Alex-3}, machine learning (ML) has emerged as a transformative tool. ML accelerates the design process by predicting material properties with high accuracy~\cite{CGCNN,MEGNet,ALIGNN}, and hence narrows down the candidate materials before costly DFT computations are performed. 
However, this combined ML-DFT approach has its limitations. The all-important question is how the initial pool of candidate materials 
is constructed. Most often, this is achieved by modifying existing materials for which there are no universally guiding principles or systematic rules. Hence, it is limited by prior human knowledge and human
intuition. This limits the scalability and innovation potential of current material design methodologies.   

Generative machine learning models have recently revolutionized synthetic data generation in fields such as image synthesis~\cite{image-gen-1,image-gen-2}, text generation~\cite{PaLM,ChatGPT}, and video creation~\cite{video-gen-1,video-gen-2}. They have incredible capabilities to uncover patterns by learning underlying data distribution. In the context of  materials design, variational autoencoders (VAEs)~\cite{VAE1,VAE2,iMatGen,FTCP,CDVAE,ACS-JCIM,Court-ChemMat} and generative adversarial networks (GANs)~\cite{GAN,DCGAN,CubicGAN,ZeoGAN,CrystalGAN,CCCG} have been successfully applied to design novel materials for diverse applications~\cite{CSFE,superconductor,Lyngby2022,MagGen}. 

Although VAEs and GANs seem promising for generative modeling, they have inherent challenges that can limit their effectiveness in materials design. VAEs often suffer from posterior collapse~\cite{posterior-collapse}, where the latent space fails to capture meaningful structural or chemical variations, leading to blurred or unrealistic material representations. On the other hand, GANs, while capable of generating sharper structures, are prone to mode collapse~\cite{mode-collapse}, where they fail to explore the full diversity of possible materials, often missing key compositions or phases. Additionally, their adversarial training process introduces instability, making it difficult to ensure reliable generation of physically meaningful materials. These challenges limit the applicability of VAEs and GANs in exploring vast chemical spaces and designing materials with targeted functionalities, underscoring the need for more robust generative approaches~\cite{dhariwal2021diffusionmodelsbeatgans}.

Diffusion models~\cite{sohldickstein,DDPM,song,EDM}, a recent breakthrough in generative machine learning, overcome these limitations. By modelling the data generation process as a gradual transformation from noise to structure, diffusion models achieve higher-quality and more diverse samples. Their superior performance positions them as the next-generation tool for efficient materials design. 

Recent advances in diffusion-based generative models~\cite{mattergen,DiffCSP,DiffCSP++,WyckoffDiff,walsh} have shown great potential in generating chemically valid, diverse crystalline materials. However, a key limitation of many of these models is the added complexity introduced by strong physics-based priors applied separately to atom types, atomic coordinates, and lattice parameters~\cite{mattergen}. This often necessitates customized diffusion processes, such as masked diffusion for atom types or symmetric diffusion for the lattice. As a result, these models struggle to fully capture the intricate correlations and joint distribution among structural components. Instead, dependencies are typically enforced conditionally during the denoising process. It may lead to inconsistencies among components, such as chemically implausible atom-lattice combinations. Alternative frameworks like CDVAE~\cite{CDVAE} and SyMat~\cite{SyMat} learn atom types and lattice parameters separately through VAEs and employ score networks for conditional generation of atomic positions. It also suffers from potential error propagation between decoupled modules and do not capture the full joint distribution of structural components. DiffCSP~\cite{DiffCSP}, on the other hand, focuses solely on structure prediction, assuming atom types are given, limiting its generative capability.

Motivated by the success of diffusion models in learning complex visual structures from images without explicit priors, we explore whether a purely data-driven approach can similarly capture the intricate correlations inherent in crystal structures. We demonstrate that, with a sufficiently large dataset, a highly expressive diffusion model can implicitly learn crystallographic priors directly from data, eliminating the need for handcrafted constraints. 

In this study, we introduce DiffCrysGen, a score-based diffusion model for generating diverse inorganic crystalline materials. Our model is trained on a 2D point cloud representation of all structural components of materials, and learns their joint distribution through a single score network. Rather than enforcing explicit priors, it adopts a purely data-driven approach. It simultaneously generates atom types, fractional coordinates, and lattice parameters in an end-to-end diffusion process. By learning dependencies directly rather than imposing them through conditional structures, DiffCrysGen simplifies the overall architecture, reduces computational cost, avoids error propagation across decoupled modules, and establishes a new paradigm for generative materials design. As a proof of concept, we leverage DiffCrysGen to design rare-earth-free magnetic materials with high saturation magnetization ($M_s$), targeting sustainable alternatives to conventional permanent magnets. 

\section{Methods}
\subsection{Representation of Crystalline Materials}

A crystalline material is characterized by its unit cell, the smallest repeating structural motif that defines its entire structure. To specify the unit cell uniquely, three fundamental components are required: the lattice, the atomic composition, and the atomic positions within the unit cell. The lattice determines the unit cell's shape through lattice constants ($a, b, c$) and angles ($\alpha, \beta, \gamma$). The atomic composition specifies the chemical elements present, while their spatial arrangement is defined by their positions within the unit cell.

We employ the Invertible Real-Space Crystallographic Representation (IRCR), introduced in our earlier work \cite{MagGen}, to systematically encode the unit cell. IRCR consists of five matrices, each capturing a distinct aspect of the crystalline material, as summarized in Table~\ref{tab:IRCR}. While the property matrix is not essential for unconditional material generation, it plays a crucial role in property-conditioned material design and in training predictive models using IRCR.

This representation ensures invertibility, allowing for the complete reconstruction of a material’s lattice, atomic positions, and composition from its encoded data. Additionally, IRCR is flexible enough to accommodate materials with diverse compositions and structures. In this work, we have used it to encode up to ternary materials, but it can be readily extended beyond that.

\begin{table}[t]
    \centering
    \caption{Components of the Invertible Real-Space Crystallographic Representation (IRCR).}
    \renewcommand{\arraystretch}{1.5} 
    \begin{tabular}{|c|c|p{3.8cm}|}
        \hline
        \textbf{Component} & \textbf{Shape} & \textbf{Description} \\
        \hline
        $E$ (Element Matrix) & $94 \times 3$ & One-hot encoding of chemical elements (up to $Z = 94$). \\
        $L$ (Lattice Matrix) & $2 \times 3$ & Lattice constants ($a, b, c$) and angles ($\alpha, \beta, \gamma$). \\
        $C$ (Coordinate Matrix) & $n_{\text{sites}} \times 3$ & Fractional atomic coordinates in the unit cell. \\
        $O$ (Occupancy Matrix) & $n_{\text{sites}} \times 3$ & One-hot encoding of site occupancy. \\
        $P$ (Property Matrix) & $6 \times 3$ & Elemental properties, including atomic number ($Z$), electronegativity ($e$), period number ($p$), group number ($g$), valence electron number ($n_v$), and atomic fraction ($s$). \\
        \hline
    \end{tabular}
    \label{tab:IRCR}
\end{table}


\subsection{Diffusion Model}
The essence of generative machine learning lies in modeling the underlying probability distribution of a dataset. Once this distribution is captured, it becomes possible to generate new samples that share similar characteristics with the original data. Here we implement a score-based diffusion model to learn this distribution through a stochastic differential equation (SDE)-based framework.

Let the original data be denoted by $\bm{x_0}$, characterized by the unknown complex distribution $p_{\text{data}}(\bm{x_0})$. This data represents the IRCRs of crystalline materials. In the forward diffusion process, indexed by a continuous time variable $t \in [0,T]$, we smoothly transform $p_{\text{data}}(\bm{x_0})$ to a known prior distribution $p_T(\bm{x_T})$ by slowly injecting noise. It can be modelled by the following SDE :

\begin{equation}
    d\bm{x} = f(\bm{x},t)dt + g(t)d\bm{w},
\end{equation}

\noindent where $f(\bm{x},t)$ and $g(t)$ are the drift and diffusion coefficients, respectively. $\bm{w}$ represents the standard Wiener process (a.k.a. Brownian motion). Usually the drift term is of the form $f(\bm{x},t)=f(t)\bm{x}$. The conditional distribution of $\bm{x_t}$ given $\bm{x_0}$, also known as the perturbation kernel of the SDE, can be formulated as a Gausssian distribution:  
\begin{equation}
   p(\bm{x_t}|\bm{x_0}) = \mathcal{N}(\bm{x_t};s(t)\bm{x_0},s^2(t)\sigma^2(t)I),
\end{equation}
where $\bm{x_t}$ is the noisy data at time t. 
The mean of this distribution is a scaled version of the original data, while the variance 
controls the noise level added to the data.

The scaling factor $s(t)$ and the noise level $\sigma(t)$ can be expressed in terms of the drift and diffusion coefficients of the SDE:
\begin{equation}
  s(t)=exp(\int_0^tf(\xi)d\xi ) 
\end{equation}

\begin{equation}
  \sigma(t)=\sqrt{\int_0^t \frac{g^2(\xi)}{s^2(\xi)} d\xi }
\end{equation}

We implemented a variance-exploding (VE) diffusion process where the noise level $\sigma(t)$ increases linearly with time. For simplicity, we set $s(t)=1$, ensuring that noise is added directly to the data without scaling the original data $\bm{x_0}$. The noisy data $\bm{x_t}$ at any time $t$ is then obtained as:
\begin{equation}
  \bm{x_t} = \bm{x_0} + \sigma(t)\bm{\epsilon},
\end{equation}
where $\bm{\epsilon}\sim\mathcal{N}(0,I)$ represents noise drawn from a standard normal distribution. In this setup, the forward diffusion process is thus governed entirely by the stochastic noise, with a zero drift term. The data is diffused with the true data $\bm{x_0}$ as the mean, while the variance grows with time. At the final step $T$, the data structure is fully destroyed and transformed into pure noise with a maximum variance $\sigma_{max}$, represented by $p_T$, which contains no information about the original data distribution $p_{data}$.

In the reverse diffusion process, starting from $\bm{x_T} \sim p_T$, new samples can be generated by reversing the forward SDE. Like the forward process, the reverse process is also a diffusion process, but it evolves backward in time. The reverse-time SDE is given by:

\begin{equation}
    d\bm{x} = [f(\bm{x},t)-g^2(t)\nabla_{\bm{x}} \log p_t(\bm{x})]dt + g(t)d\bm{\bar{w}},
    \label{reverse-SDE}
\end{equation}
where $\bar{\bm{w}}$ is the Wiener process corresponding to time flowing backward from $T$ to $0$, and $dt$ represents an infinitesimal negative timestep. Here, $p_t(\bm{x})$ denotes the marginal distribution at time $t$, and $\nabla_{\bm{x}} \log p_t(\bm{x})$ is known as the score function, which captures the gradient of the log probability. It points towards higher density of data at any given noise level. 

For our VE setup, the reverse SDE Eq.(\ref{reverse-SDE}) becomes 

\begin{equation}
    d\bm{x} = -2\dot{\sigma}(t)\sigma(t)\nabla_{\bm{x}} \log p_t(\bm{x}) dt + \sqrt{2\dot{\sigma}(t)\sigma(t)}d\bm{\bar{w}}.
    \label{VE-setup}
\end{equation}

Here, the only unknown term is the score function of the marginal distribution. By accurately estimating the score function, we can numerically solve Eq.(\ref{VE-setup}), generating new data by simulating the process backward to $t=0$.

To estimate the score of the marginal distribution, we train a denoiser function $D_\theta(\bm{x_t};\sigma(t))$, implemented as a noise-conditional neural network.
This network takes the noisy data $\bm{x_t}$ and the corresponding noise level $\sigma(t)$ as inputs and predicts the clean data $\bm{x_0}$.

The training of the denoiser is guided by minimizing the following loss function: 
\begin{equation}
    \mathcal{L} = 
    \mathbb{E}_{\bm{x_0}}
    \mathbb{E}_{\bm{x_t}}
    \mathbb{E}_{\sigma}
    [ \lambda(\sigma)
||D_\theta(\bm{x_t};\sigma(t)) - \bm{x_0}||^2_2
    ],
    \label{loss}
\end{equation}
where $\bm{x_0}$, $\bm{x_t}$ and $\sigma$ are sampled from $p_{data}$, $p(\bm{x_t}|\bm{x_0})$ and $p_{train}$, respectively. Here $p_{train}$ specifies the noise level distribution used for training. The weighting function $\lambda(\sigma)$ regulates the contribution of different noise levels during training, ensuring robustness across a wide range of noise levels. After training, the score function $\nabla_{\bm{x}} \log p_t(\bm{x_t})$  can be computed directly from the denoiser using the relation:
\begin{equation} 
\nabla_{\bm{x}_t} \log p_t(\bm{x}_t) =
\frac{D_{\theta}(\bm{x}_t, \sigma(t)) - \bm{x}_t}
{\sigma^2(t)}
\end{equation}
With the score function in hand, new samples can be generated by solving the reverse SDE Eq.(\ref{VE-setup}).
This approach leverages the denoiser to approximate the score function at each time step, facilitating an efficient and accurate transition from the final noisy distribution $p_T(\bm{x_T})$ back to the data distribution $p_{data}(\bm{x_0})$.

\section{Results and Discussion}
\subsection{Data Set Construction}

To train the diffusion model, we curated a large and diverse dataset from the Alexandria~\cite{Alex-1,Alex-2,Alex-3} database, which contains DFT-computed material properties. The original database comprises 4,489,295 materials. We first filtered structures with $\le20$ atoms per unit cell, elementary, binary, or ternary compositions, and elements up to Pu (Z=94), reducing the dataset to 2,584,689 materials.

Next we applied constraints on lattice parameters ($\le25$\AA), convex hull energy ($E_{hull}\le0.1$ eV/atom), and formation energy ($h_{form}\le0$ eV/atom), yielding 537,049 materials. Since this set included many non-magnetic materials, we further filtered for structures with a saturation magnetization $M_s\ge10^{-5}$ Tesla, resulting in 155,534 materials. We call this curated dataset Alex-1.

\begin{figure*}
    \centering
    \includegraphics[scale=0.40]{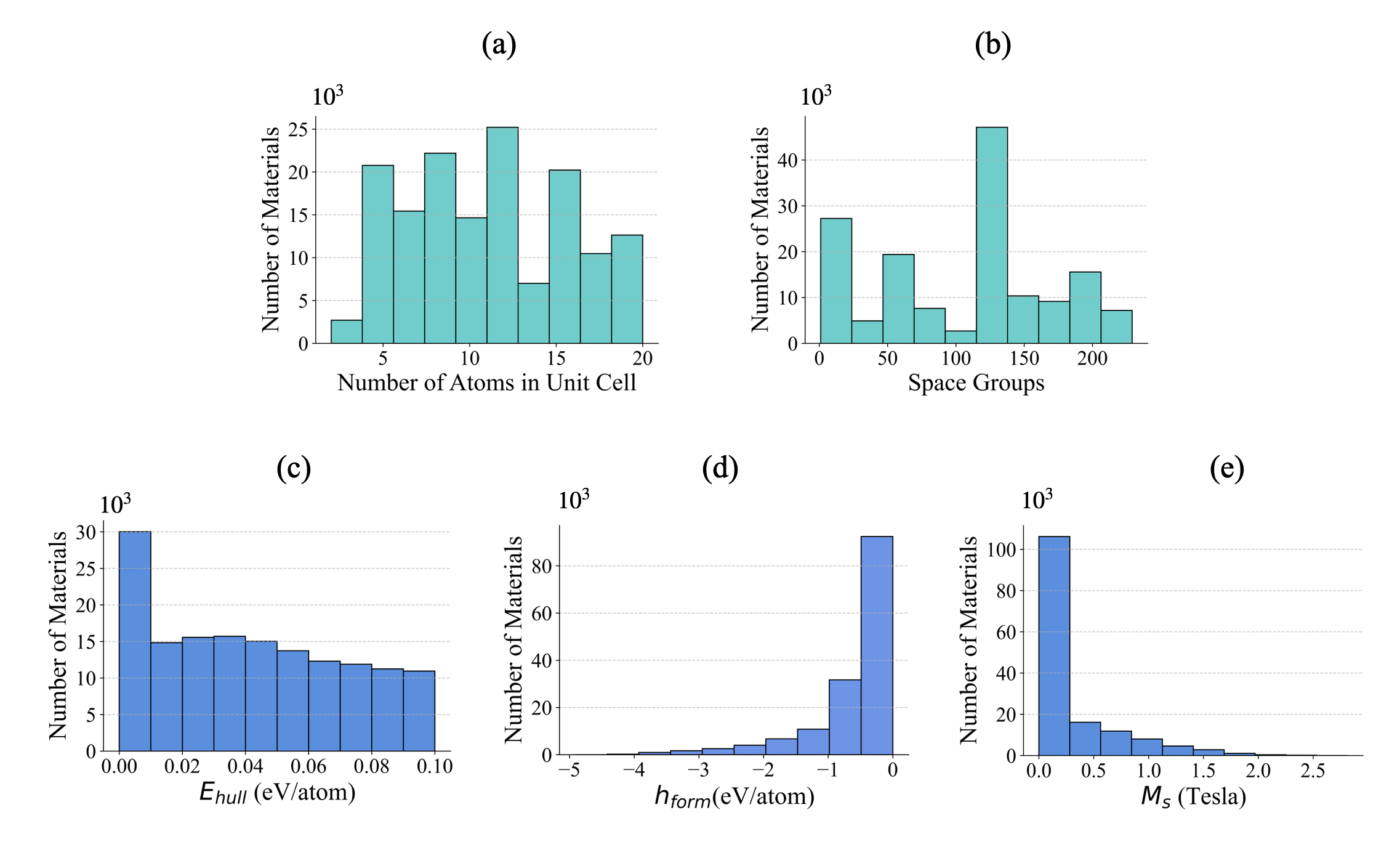}
    \caption{Distribution of (a) number of atoms in unit cell, (b) space groups, (c) convex hull, (d) formation energy and (e) saturation magnetization of materials in the Alex-1 dataset.}
    \label{training-data}
\end{figure*}

The majority of the materials ($94\%$) in Alex-1 are ternary.
Fig.~\ref{training-data}(a) shows the distribution of atoms per unit cell in the dataset, revealing that most materials contain more than 10 atoms in the unit cell.
In Fig.~\ref{training-data}(b) we plot the distribution of space groups, highlighting that the dataset spans all crystal classes. The majority of the materials belong to tetragonal($38.8\%$), orthorhombic($20.2\%$), and monoclinic($15.9\%$) crystal systems. 

Fig.~\ref{training-data}(c)-(d) illustrate the distributions of $E_{hull}$ and $h_{form}$. $8.8\%$ of the materials lie directly on the convex hull. We have used stable and metastable materials close to the hull to enhance the model's ability to generate energetically viable structures. 

In Fig.~\ref{training-data}(e) we show the distribution of saturation magnetization ($M_s$). Most materials exhibit relatively low magnetization. Notably, only 21\% of the materials have a saturation magnetization exceeding 0.5 Tesla, while those with exceptionally high magnetization constitute a small fraction. This underscores the inherent challenge in designing high-performance magnetic materials.

Finally, we split Alex-1 into an 80:10:10 ratio for training, validation, and testing the diffusion model. Fig.~\ref{parity-plot}(a) presents the learning curve, which exhibits stable convergence, with the loss decreasing smoothly and gradually, without abrupt fluctuations, indicating a well-behaved and effective training process.

\subsection{Property Prediction Model}

We trained two separate property prediction models, built using convolutional neural networks~\cite{CNN}, with IRCR as the input feature: one for $h_{form}$ and another for $M_s$. For $M_s$ prediction, we used the Alex-1 dataset. However, since Alex-1 contains materials with negative $h_{form}$ only, we expanded the dataset by including unstable materials also to improve the robustness of $h_{form}$ predictor. Specifically, we augmented Alex-1 with 309,430 unstable materials, resulting in a combined dataset of 464,964 materials, which we refer to as Alex-2. This expanded dataset was used exclusively for training the $h_{form}$ regressor. 

\begin{figure*}
    \centering
    \includegraphics[scale=0.4]{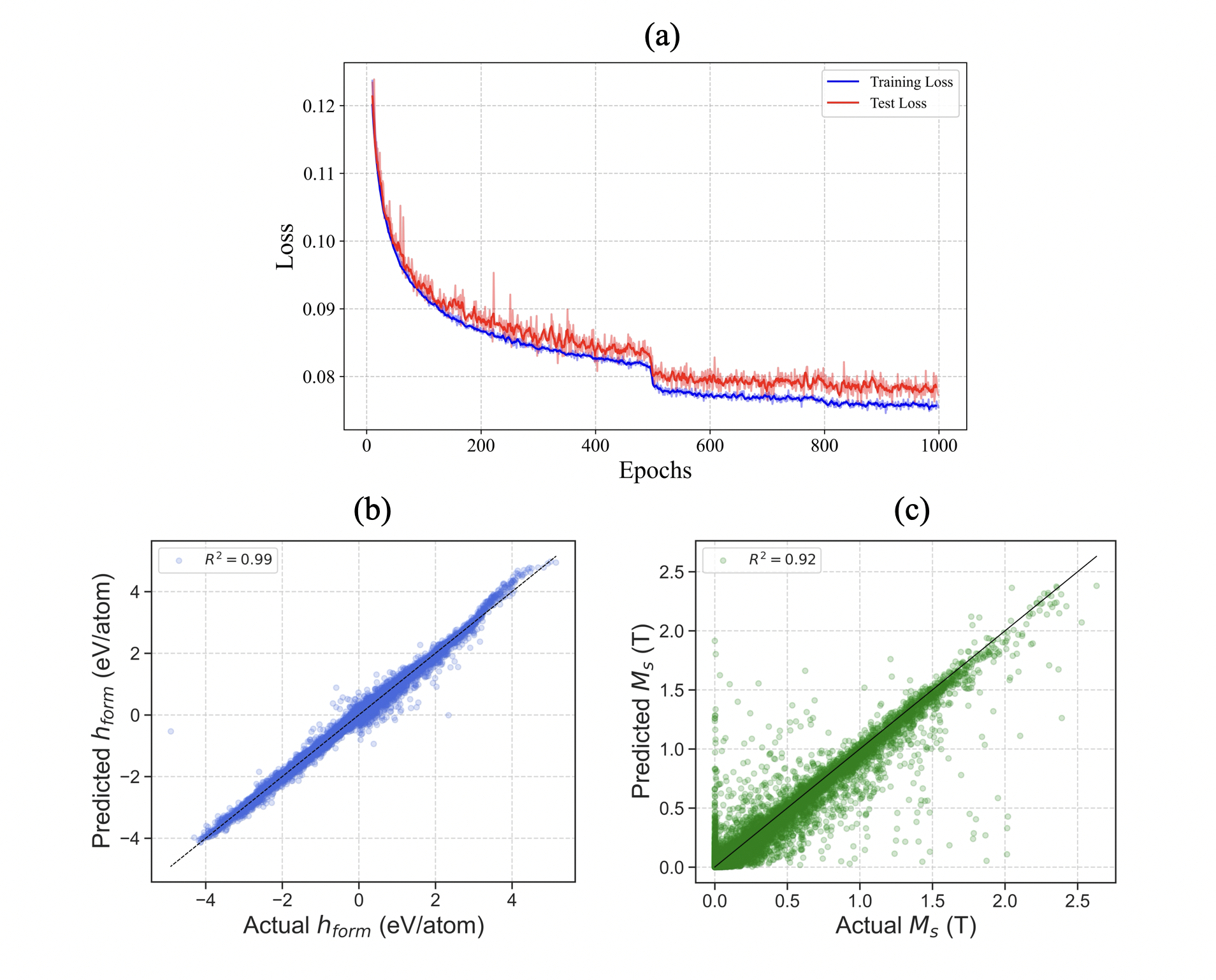}
    \caption{(a) Learing curve of the diffusion model, (b)-(c) Parity plots of $h_{form}$ and $M_s$ predictions in the test set.}
    \label{parity-plot}
\end{figure*}

The property prediction models exhibit exceptional accuracy, with the formation energy ($h_{form}$) predictor achieving a mean absolute error (MAE) of 0.046 eV/atom and an $R^2$ score of 0.99, while the saturation magnetization ($M_s$) predictor attains an MAE of 0.054 Tesla with an $R^2$ score of 0.92. In Fig.~\ref{parity-plot}(b)-(c), we present the parity plots for $h_{form}$ and $M_s$, comparing actual and predicted values. The predictions exhibit a well-balanced distribution across the entire property range, demonstrating the robustness of both models. 

While state-of-the-art graph neural networks trained on the entire Alexandria database~\cite{SCHMIDT2024101560} have achieved lower MAEs (0.016 eV/atom for $h_{form}$ and 0.031 Tesla for $M_s$), our models demonstrate comparable performance even with a much smaller training data. 
This highlights the efficiency of our model, making it a powerful tool for high-throughput screening of novel magnetic materials.

\subsection{Generated Materials}



Using our trained diffusion model, we generated a total of 1,264,466 candidate materials, among which 959,122 represent novel compositions absent from the training set. To align with our objective of discovering rare-earth-free magnetic materials, we excluded compositions containing rare-earth elements by restricting the chemical space to elements with atomic numbers $Z\le54$, yielding a refined set of 228,009 compositions. To assess the chemical diversity of this rare-earth-free subset, we first analyzed its compositional complexity: 24.16\% of the materials are binary, while the remaining 75.84\% are ternary. The dataset spans a wide range of chemistry, including transition-metal alloys (41.51\%), oxides (5.93\%), halides (13.13\%), and chalcogenides (7.19\%), highlighting the model’s ability to generate chemically diverse compounds.

Notably, $28.19\%$ of these materials have a space group greater than 16, indicating that they belong to high-symmetry crystal classes, excluding triclinic and monoclinic structures. This represents a significant improvement over previous VAE or GAN-based models, which predominantly generated triclinic materials (more than $99\%$ of the time) as highlighted elsewhere~\cite{PGCGM}.   

In our previous work, where the VAE~\cite{MagGen} model was trained on a dataset of 40,048 materials, $99.99\%$ of the generated materials were either monoclinic or triclinic, with $94.49\%$ belonging to space group P1. To examine whether a larger dataset improves performance, we trained the same VAE architecture on Alex-1 that was used to train the diffusion model. However, the results remained nearly unchanged: $99.82\%$ of the generated materials were monoclinic or triclinic, and $94.10\%$ had space group P1.

In contrast, diffusion model produced $71.81\%$ monoclinic and triclinic materials, with only $51.43\%$ belonging to space group P1. This demonstrates that the diffusion model inherently learns crystal symmetry directly from raw data without explicit symmetry-aware training.

In the image domain, diffusion models have already been shown to generate higher-quality images compared to VAEs and GANs. In materials science, we hypothesize that this advantage extends to structural symmetry, as reflected in the space group distributions of the generated materials. This highlights the diffusion model's ability to capture the underlying structural symmetries of materials more effectively than traditional generative approaches.

Although promising, the diffusion model still exhibits a bias toward lower-symmetry structures compared to the training dataset, where only $17.97\%$ of materials belong to monoclinic and triclinic classes. This suggests that the model does not fully capture the distribution of crystal symmetries present in the training data.

However, recent studies~\cite{DiffCSP,DiffCSP++,WyckoffDiff} have shown that incorporating symmetry-aware training can improve the generation of high-symmetry materials. We believe that applying such techniques to our diffusion model could enhance its ability to produce materials with higher space groups more efficiently.

These materials were further evaluated using the trained property prediction models for formation energy ($h_{form}$) and saturation magnetization ($M_s$).

Of the filtered set, 164,716 materials exhibited negative formation energy and 104,574 materials had $h_{form}\le-0.2$ eV/atom. Among the latter, 2,416 materials displayed $M_s\ge1$T, a requirement for permanent magnets to generate stronger fields. Since performing DFT on all 2,416 materials is time-consuming, we further pruned the data set by selecting materials with a minimum interatomic distance ($d_{\rm min}\ge1\AA$), yielding 862 materials. This can also be rationalized from the fact that most atom-atom distances in materials
would be greater than 1~\AA.

Within this final set, 635 materials are ternary. We focus on ternary materials only because they offer a diverse range of structural and compositional possibilities, increasing the likelihood of discovering novel materials. Additionally, 140 materials belong to space groups larger than 16, selected to focus on structures with higher symmetry. In our earlier work~\cite{JMMM}, we conducted a statistical analysis of a data set for permanent magnets sourced from novomag~\cite{novomag} and Novamag~\cite{novamag}. Our findings indicated that orthorhombic, tetragonal, and hexagonal materials exhibit strong uniaxial magnetocrystalline anisotropy. Consequently, we prioritized high-symmetry materials. These 140 candidates are then subjected to density functional theory (DFT) calculations to identify rare-earth-free permanent magnet candidates.

\subsection{DFT Validation}
To evaluate the performance of diffusion model and property predictors in generating new materials that meet the desired properties, we employ the following two metrics:

\begin{enumerate}
    \item \textbf{Validity rate}: The fraction of generated materials for which structural relaxation is successfully completed in DFT.
    \item \textbf{Success rate}: The fraction of generated materials that satisfy the design targets after structure optimization, as determined through DFT calculations.
\end{enumerate}

Out of the 140 candidate materials, structure optimization was successful for 121 materials, while optimization failed to converge to any local minima for the remaining 19 materials, yielding a validity rate of $86.42\%$. Notably, while all pre-optimized structures belonged to space groups greater than 16, many transitioned to lower-symmetry monoclinic or triclinic phases after optimization. Specifically, 46 of the 121 optimized structures underwent symmetry reduction having space group below 16, highlighting the importance of structural relaxation in accurately determining crystal symmetries. This finding underscores the inherent tendency of some predicted structures to stabilize in lower-symmetry configurations, which could impact their physical and electronic properties.

Among the 121 valid materials, 111 exhibit negative formation energy, and 107 materials meet the design target of formation energy ($h_{form}\le-0.2$ eV/atom). Furthermore, 107 materials satisfy the design target of magnetization ($M_s\ge1$ T). Altogether, 97 materials fulfill both the design targets. This corresponds to a success rate of $80.16\%$ among structurally valid materials, and an overall success rate of $65\%$. This represents a significant achievement, as it demonstrates our ability to design materials within a highly constrained region of the vast material space. For context, only $5.78\%$ of materials in the training set satisfy both design targets. The observed success rate highlights the efficacy of our approach, showcasing its potential to navigate and uncover rare, high-performing candidates in complex chemical spaces.  

To rigorously assess thermodynamic stability beyond the formation enthalpy ($h_{form}$), we computed the convex hull energy ($E_{hull}$) for all 97 materials that satisfied both design targets. Among them, 54 materials exhibit $E_{hull} \leq 0.4$ eV/atom, including 13 materials within 0.1 eV/atom of the convex hull.

A viable permanent magnet must combine high saturation magnetization with strong uniaxial magnetocrystalline anisotropy energy ($K_1$), which stabilizes magnetization through spin-orbit coupling. To identify such candidates, we focused on materials satisfying both design targets with $E_{hull} \leq 0.4$ eV/atom, acknowledging the inclusion of metastable phases. This broader criterion aims to capture potential high-anisotropy materials that might be overlooked if limited to phases near the convex hull. While non-equilibrium synthesis techniques may be required for synthesizing such metastable phases, this approach ensures a more comprehensive exploration of promising candidates. 

We then computed $K_1$ for all these 54 materials. While three exhibited easy-plane anisotropy, the remaining 51 displayed the essential uniaxial anisotropy required for permanent magnet applications. Among these, 17 materials demonstrated significant anisotropy ($K_1 \geq 0.5$ MJ/m³), with 10 exhibiting very strong anisotropy ($K_1 \geq 1$ MJ/m³), meeting the benchmark for high-performance permanent magnets.

For a comprehensive stability assessment, we performed phonon dispersion calculations to evaluate dynamical stability. We first analyzed the 10 materials with high magnetic anisotropy ($K_1 \geq 1$ MJ/m³). Additionally, we also examined 13 materials closest to the convex hull ($E_{hull} \leq 0.1$ eV/atom), irrespective of their $K_1$ value, since they are more likely to be experimentally synthesizable. While these materials may not exhibit high anisotropy, they remain promising candidates for further modifications or practical applications.

Among the high-anisotropy candidates, 5 were dynamically stable, showing no imaginary phonon modes. 8 materials near the convex hull also exhibited dynamical stability, bringing the total number of dynamically stable materials to 13. All these materials are summarized in Table~\ref{DFT}.

\begin{figure*}
    \centering
    \includegraphics[scale=0.40]{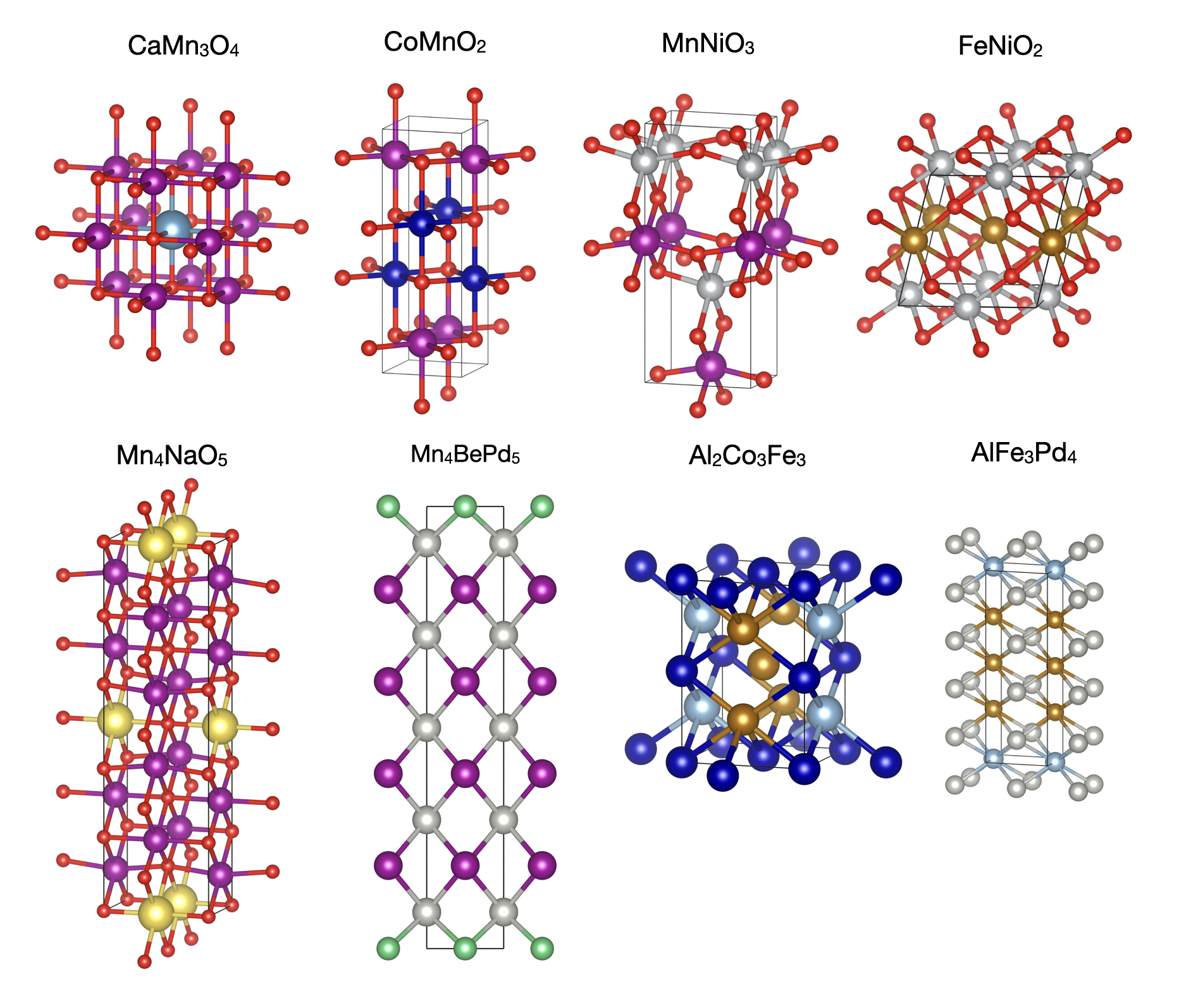}
    \caption{Visualization of the crystal structures of the top eight materials that are closest to the convex hull ($E_{hull} \leq 0.1$ eV/atom), dynamically stable, and exhibit high saturation magnetization ($M_s \geq 1$ T).}
    \label{crystal-structure-1}
\end{figure*}

\begin{figure*}
    \centering
    \includegraphics[scale=0.40]{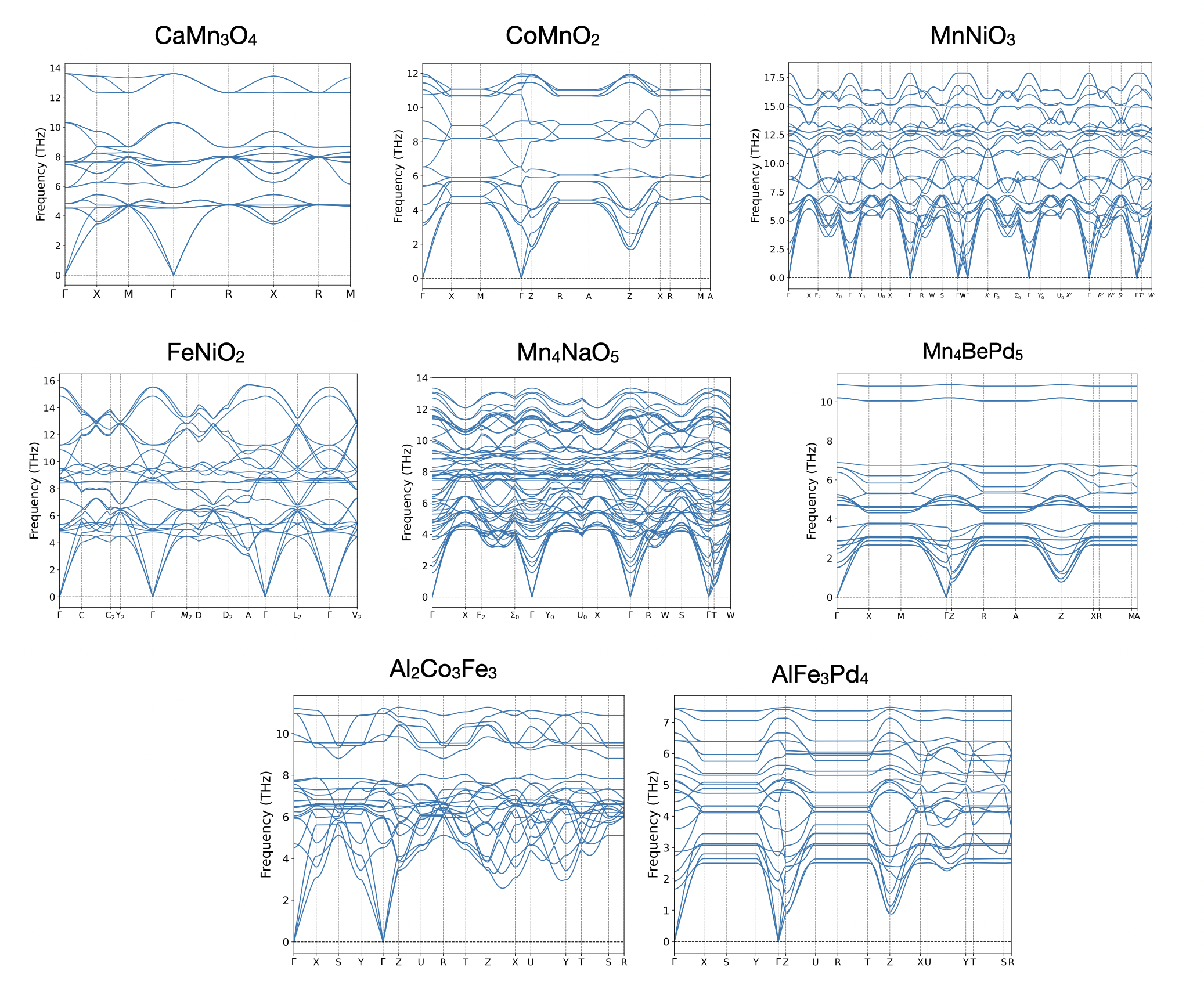}
    \caption{Phonon dispersions of the top eight materials that are closest to the convex hull ($E_{hull} \leq 0.1$ eV/atom), dynamically stable, and exhibit high saturation magnetization ($M_s \geq 1$ T).}
    \label{phonon-1}
\end{figure*}

The top eight materials are the ones closest to the hull ($E_{hull}\le0.1$ eV/atom). The crystal structures are visualized in Fig.~\ref{crystal-structure-1}, with their phonon dispersions shown in Fig.~\ref{phonon-1}.
These materials exhibit a diverse range of compositions, including transition metal oxides ($\text{Ca}\text{Mn}_3\text{O}_4$, $\text{Co}\text{Mn}\text{O}_2$, 
$\text{Mn}\text{Ni}\text{O}_3$,
$\text{Fe}\text{Ni}\text{O}_2$,
), alkali-metal-containing oxide ($\text{Mn}_4\text{Na}\text{O}_5$), and intermetallic compounds ($\text{Mn}_4\text{Be}\text{Pd}_5$, $\text{Al}_2\text{Co}_3\text{Fe}_3$, $\text{Al}\text{Fe}_3\text{Pd}_4$).
Structurally, they span high-symmetry crystal classes such as cubic, tetragonal, and orthorhombic, with one in monoclinic phase. Their proximity to the convex hull suggests that these materials are thermodynamically favorable and more likely to be experimentally realizable under normal conditions. While their magnetic anisotropy is relatively low, their high saturation magnetization ($M_s \geq 1$ T) indicates potential applications in soft magnetic materials. Moreover, their stability and compositional diversity offer opportunities for further optimization, such as strain engineering or chemical doping, to enhance magnetic anisotropy.

\begin{figure*}
    \centering
    \includegraphics[scale=0.40]{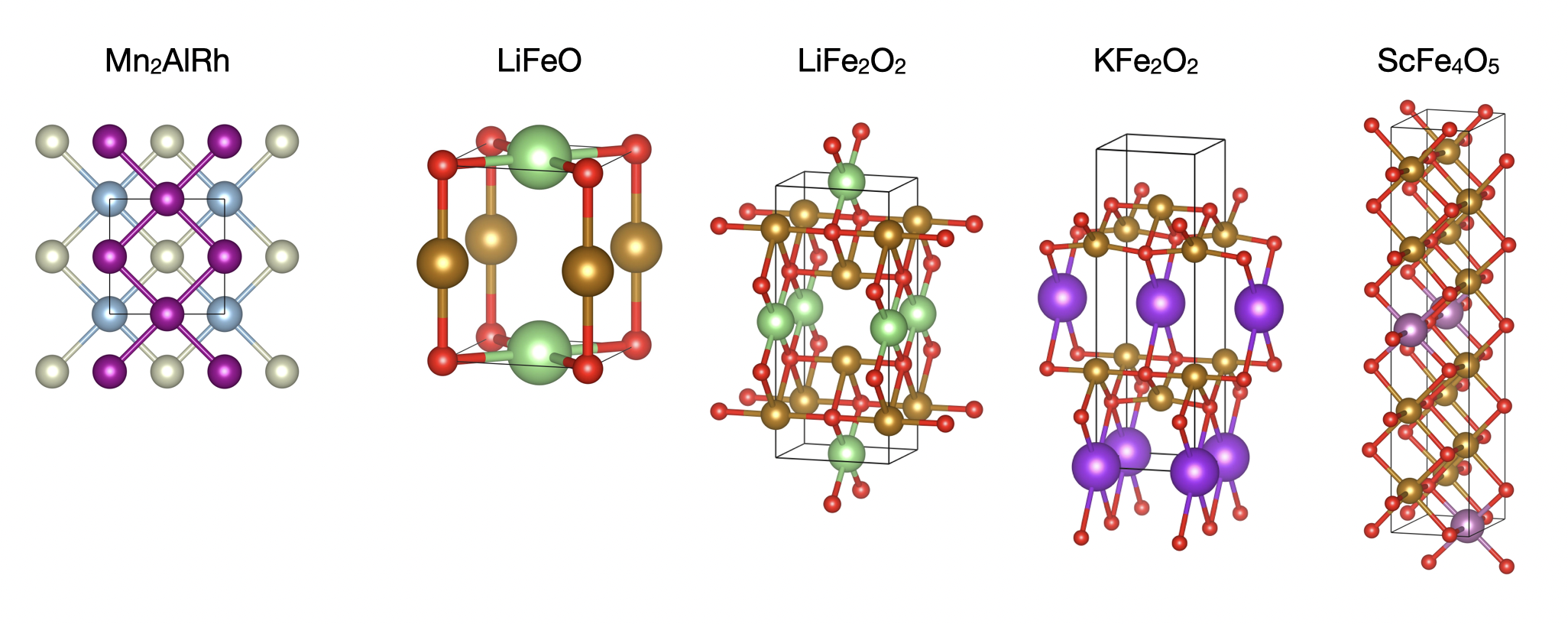}
    \caption{Visualization of the crystal structures of the bottom five materials, which exhibit high saturation magnetization ($M_s \geq 1$ T) with large magnetocrystalline anisotropy ($K_1\ge1$ MJ/m$^3$), and also dynamically stable.}
    \label{crystal-structure-2}
\end{figure*}

\begin{figure*}
    \centering
    \includegraphics[scale=0.45]{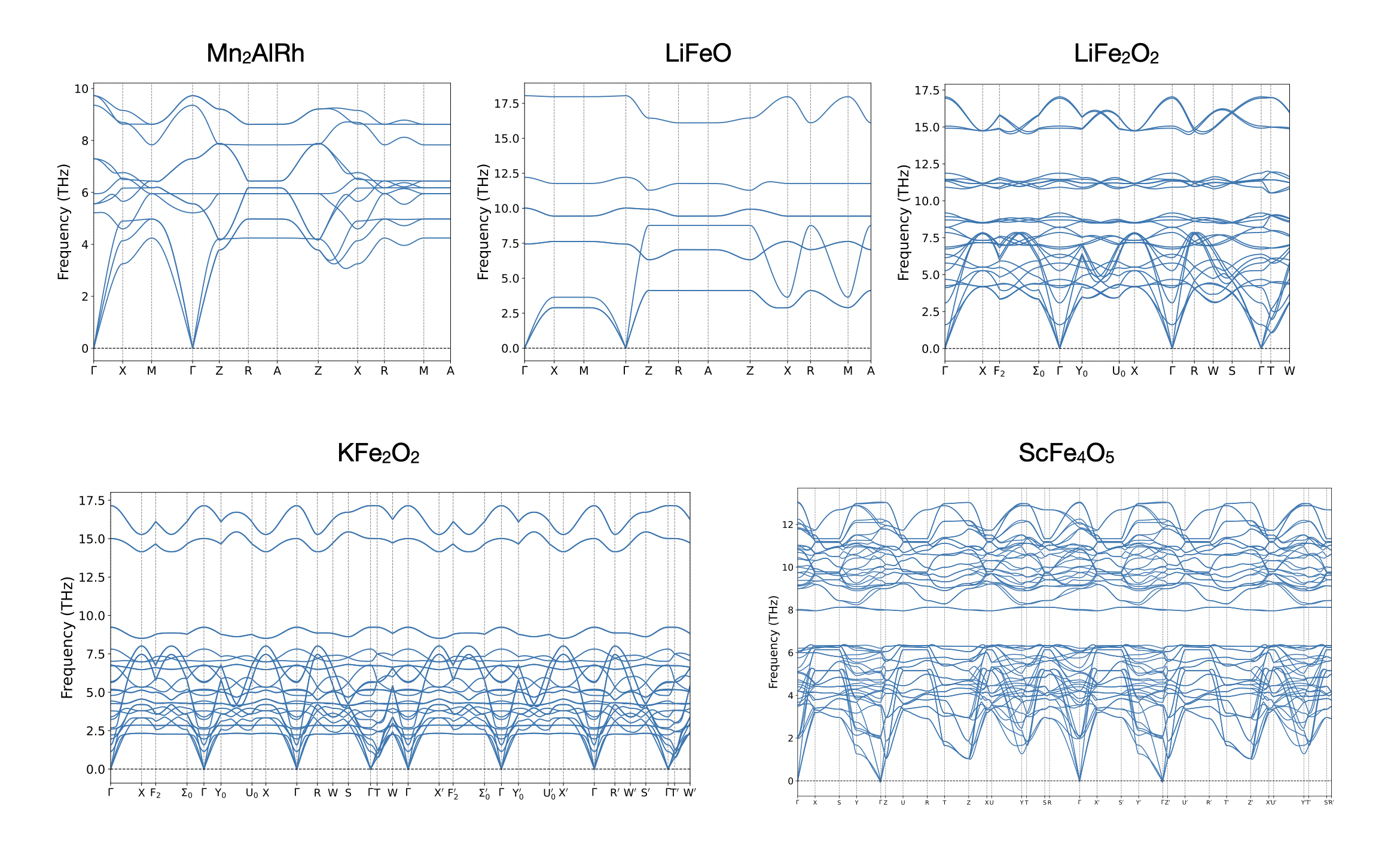}
    \caption{Phonon dispersions of the bottom five materials, which exhibit high saturation magnetization ($M_s \geq 1$ T) with large magnetocrystalline anisotropy ($K_1\ge1$ MJ/m$^3$), and also dynamically stable.}
    \label{phonon-2}
\end{figure*}

The bottom five materials, characterized by high magnetic anisotropy ($K_1\ge1$ MJ/m$^3$), includes $\text{Mn}_2\text{Al}\text{Rh}$, $\text{LiFeO}$, $\text{Li}\text{Fe}_2\text{O}_2$, $\text{K}\text{Fe}_2\text{O}_2$, and $\text{Sc}\text{Fe}_4\text{O}_5$, all belonging to tetragonal and orthorhombic crystal classes. Their crystal structures are visualized in Fig.~\ref{crystal-structure-2}, with phonon dispersions shown in Fig.~\ref{phonon-2}.  These materials not only exhibit high saturation magnetization but also possess significantly larger $K_1$ values, making them promising candidates for rare-earth free permanent magnet applications. Notably, $\text{LiFeO}$, $\text{Li}\text{Fe}_2\text{O}_2$ and $\text{K}\text{Fe}_2\text{O}_2$ exhibit exceptionally large $K_1$ values exceeding 4 MJ/m$^3$, comparable to the widely used rare-earth magnet $\text{Nd}_2\text{Fe}_{14}\text{B}$.
While their $E_{hull}$ values are slightly higher than those of the first set, the combination of strong magnetization and large anistropy makes them attractive targets for further theoretical and experimental exploration.

\begin{table*}[ht]
    \centering
     \caption{DFT computed properties of the final 14 materials, generated by the diffusion model, which satisfy the design targets and are dynamically stable with $E_{hull}$ up to 0.4 eV/atom.   
     }
     \vspace{0.2cm}
    \begin{tabular}{|c|c|c|c|c|c|c|}
    \hline
    Composition & Crystal Type & Space Group & $h_{form}$ (eV/atom) & $M_s$(T) & $E_{hull}$ (eV/atom) & $K_1$ (MJ/m$^3$) \\
    \hline
    $\text{Ca}\text{Mn}_3\text{O}_4$ & Cubic & Pm-3m (221) & -2.33 & 1.81 & 0.02 & 0 \\ 
    $\text{Co}\text{Mn}\text{O}_2$ & Tetragonal & P4/nmn (129) & -1.587 & 2.18 & 0.046 & 0.22 \\ 
    $\text{Mn}\text{Ni}\text{O}_3$ & Orthorhombic & Imm2 (44) & -1.638 & 1.118 & 0.09 & -0.237 \\ 
    $\text{Fe}\text{Ni}\text{O}_2$ & Monoclinic & C2/m (12) & -1.514 & 1.753 & 0.007 & 0.14 \\
    $\text{Mn}_4\text{Na}\text{O}_5$ & Orthorhombic & Immm (71) & -2.018 & 1.97 & 0.025 & 0.378 \\
    $\text{Mn}_4\text{Be}\text{Pd}_5$ & Tetragonal & P4/mmm (123) & -0.264 & 1.412 & 0.014 & 0.1 \\ 
    $\text{Al}_2\text{Co}_3\text{Fe}_3$ & Orthorhombic & Pmmm (47) & -0.281 & 1.167 & 0.048 & 0.55 \\
    $\text{Al}\text{Fe}_3\text{Pd}_4$ & Orthorhombic & Pmmm (47) & -0.261 & 1.014 & 0.091 & 0.44 \\ 
    \hline\
    $\text{Mn}_2\text{Al}\text{Rh}$ & Tetragonal & P4/mmm (123) & -0.283 & 1.18 & 0.266 & 1.625 \\
    $\text{LiFeO}$ & Tetragonal & P4/mmm (123) & -1.272 & 1.11 & 0.376 & 4.196 \\ 
    $\text{Li}\text{Fe}_2\text{O}_2$ & Orthorhombic & Immm (71) & -1.383 & 1.50 & 0.253 & 4.755 \\ 
    $\text{K}\text{Fe}_2\text{O}_2$ & Orthorhombic & Immm2 (44) & -1.177 & 0.978 & 0.267 & 4.233 \\
    $\text{Sc}\text{Fe}_4\text{O}_5$ & Orthorhombic & Pmn2$_1$ (31) & -1.941 & 1.65 & 0.198 & 1.910 \\ 
    \hline
    \end{tabular}
    \label{DFT}
\end{table*}

\subsection{Magnetic Ground State Analysis}

So far it has been implicitly assumed that all the magnetic materials have ferromagnetic ground states. This suited
our workflow because the property prediction model had been trained on magnetization data for FM only.
However, quite often the energy difference between the ferro- and anti-ferromagnetic (AFM) arrangements of spin
moments is rather small, as the strength of exchange interactions in magnetic materials $J \sim $~meV.
Therefore, for a comprehensive and correct understanding we rigorously check if the thirteen materials listed
in Table~\ref{DFT} have FM or AFM ground states.

For this, we computed the total energies of all the 13 materials in both FM and several AFM states. For each material, possible symmetry-allowed AFM arrangements were considered. Table~\ref{magnetic_gs} summarizes the energy differences between the FM and the lowest-energy AFM configuration, along with the resulting magnetic ground state.

Out of 13 materials, we found 5 materials to have FM ground state. 
Robust FM ground states were observed in materials such as $\text{Al}_2\text{Co}_3\text{Fe}_3$ and $\text{Al}\text{Fe}_3\text{Pd}_4$, both exhibiting large negative energy differences (–46.43 and –48.61 meV/atom, respectively). Among the materials with large $K_1$, $\text{LiFeO}$ and $\text{Sc}\text{Fe}_4\text{O}_5$ turn out to have FM ground state, making them the most interesting candidates for 
RE-free PMs. 

Though we parsed the generated materials to identify promising RE-free PMs, our analysis also leads to the discovery of several novel AFM candidates. Interestingly, $\text{LiFe}_2\text{O}_2$ displays a quasi-degenerate magnetic ground state, with a very small energy difference (0.25 meV/atom) between FM and AFM configurations. Such near-degeneracy could lead to tunable magnetic phases under external stimuli (e.g., electric field or strain), making it a potential multifunctional material for magnetoelectric or spintronic applications. Moreover, $\text{Mn}_2\text{Al}\text{Rh}$ and $\text{Mn}_4\text{Be}\text{Pd}_5$ exhibit strongly stabilized AFM orderings with energy differences exceeding 40 meV/atom, suggesting robust AFM interactions. These AFM materials are relevant in applications such as antiferromagnetic spintronics, where zero net magnetization and high-frequency spin dynamics are advantageous. 

\begin{table*}[ht]
\centering
\caption{Comparison of energies between ferromagnetic (FM) and antiferromagnetic (AFM) spin configurations. The magnetic ground state is determined by the lower-energy configuration. $h_{form}$ and $E_{hull}$ correspond to the ground-state magnetic ordering.}
\vspace{0.2cm}
\begin{tabular}{|c|c|c|c|c|}
\hline
Composition & $E_\text{FM}-E_\text{AFM}$ (meV/atom) & Ground State & $h_{form}$ (eV/atom) & $E_{hull}$ (eV/atom) \\
\hline
$\text{Ca}\text{Mn}_3\text{O}_4$ & 13.87 & AFM & -2.344 & 0.006 \\
$\text{Co}\text{Mn}\text{O}_2$ & 12.72 & AFM & -1.599 & 0.033 \\
$\text{Mn}\text{Ni}\text{O}_3$ & -5.25 & FM & -0.237 & 0.090 \\
$\text{Fe}\text{Ni}\text{O}_2$ & 1.01 & AFM & -1.515 & 0.005 \\
$\text{Mn}_4\text{Na}\text{O}_5$ & 8.83 & AFM & -2.027 & 0.016 \\
$\text{Mn}_4\text{Be}\text{Pd}_5$ & 45.45 & AFM & -0.309 & 0 \\
$\text{Al}_2\text{Co}_3\text{Fe}_3$ & -46.43 & FM & -0.281 & 0.048\\
$\text{Al}\text{Fe}_3\text{Pd}_4$ & -48.61 & FM & -0.261 & 0.091\\
\hline
$\text{Mn}_2\text{Al}\text{Rh}$ & 86.0 & AFM & -0.369 & 0.180\\
$\text{LiFeO}$ & -8.6 & FM & -1.272 & 0.376\\
$\text{Li}\text{Fe}_2\text{O}_2$ & 0.25 & AFM & -1.384 & 0.252\\
$\text{K}\text{Fe}_2\text{O}_2$ & 10.07 & AFM & -1.187 & 0.257\\
$\text{Sc}\text{Fe}_4\text{O}_5$ & -2.11 & FM & -1.941 & 0.198 \\
\hline
\end{tabular}
\label{magnetic_gs}
\end{table*}

\section{conclusion}

Diffusion models represent a paradigm shift in the generative design of novel materials. Our study demonstrates that, when trained on sufficiently large datasets, a powerful and expressive diffusion model can inherently learn the complex interdependence among atom types, atomic coordinates, and lattice parameters within a unified latent space. This approach obviates the need for handcrafted inductive biases or physics-based priors, thereby simplifying the generative framework.

Our model, DiffCrysGen, exemplifies this capability by successfully generating novel, stable crystalline materials aligned with targeted properties, such as high saturation magnetization in rare-earth-free compounds. While physics-based priors remain valuable, our findings suggest that deep learning models can capture essential physical principles directly from data, offering a compelling alternative pathway for materials discovery.

Although DiffCrysGen captures symmetry trends to a certain extent, enhancements are possible through more sophisticated material representations that explicitly encode crystallographic symmetries, architectural refinements, or the incorporation of larger and more diverse datasets. These avenues are the focus of our ongoing research.

In this work, we have focused on unconditional material generation. However, the foundational model can be fine-tuned using adapter modules, such as ControlNet~\cite{controlnet}, or through classifier-free-guidance~\cite{CFG} techniques to steer the generation process toward specific properties, compositions, space groups, or even specific text prompts. Such advancements will pave the way for true inverse design in materials science, a direction we are pursuing currently.

\section{Acknowledgements} 
The work was funded by the DAE, Govt. of India, through institutional funding to HRI. All calculations were performed in the cluster computing facility at HRI (https://www.hri.res.in/cluster/).

\clearpage
\section{Appendix}

\subsection{DFT Calculations}

All structure optimizations and total energy calculations were performed using spin-polarized DFT within VASP~\cite{vasp1,vasp2}. The computational parameters were chosen to be consistent with the Alexandria~\cite{Alex-1,Alex-2,Alex-3} database. We employed the projector-augmented wave (PAW)~\cite{PAW} method with Perdew–Burke–Ernzerhof (PBE)~\cite{PBE} exchange-correlation pseudopotentials and a plane-wave cutoff energy of 520 eV. Brillouin zone integrations were carried out using uniform  $\Gamma$-centered k-point meshes with a density of $0.15~\text{\AA}^{-1}$. Full structural relaxations were performed, allowing both lattice parameters and atomic positions to vary until the total energy and atomic forces converged to below $10^{-5}$ eV and $0.005$ eV/\AA, respectively. For transition metal oxides and fluorides, we employed GGA+$U$ calculations following the Dudarev approach~\cite{dudarev}, while standard GGA was used for all other compounds. The Hubbard $U$ parameters were adopted from the Materials Project (MP) database~\cite{MP}, consistent with those used in the Alexandria database (e.g., $U = 5.3$ eV for Fe).

For the calculation of the magnetocrystalline anisotropy energy constant $K_1$, we performed fully relativistic DFT calculations by including spin-orbit coupling (SOC) in the Hamiltonian. A denser \textit{k}-point mesh with a sampling density of $0.10$~\AA$^{-1}$ was used to ensure convergence of the anisotropy energy. To evaluate $K_1$, total energies were computed for magnetization oriented along the crystallographic $a$, $b$, and $c$ directions. The anisotropy constant $K_1$ was then extracted using the relation:
\begin{equation}
E = K_1 \sin^2\theta,
\end{equation}
where $E$ represents the energy cost for the spins to rotate away from a particular crystallographic direction with the lowest energy by an angle $\theta$. A positive value of $K_1$ ($K_1 > 0$) indicates easy-axis anisotropy, where the magnetization prefers to align along a specific crystallographic direction. Conversely, a negative $K_1$ ($K_1 < 0$) corresponds to easy-plane anisotropy, where the magnetization tends to lie within a plane perpendicular to the crystallographic direction.

Phonon calculations were performed using density functional perturbation theory (DFPT)~\cite{dfpt} with a strict electronic energy convergence threshold of $10^{-8}$~eV. Post-processing was performed using phonopy~\cite{phonopy1,phonopy2}.


\bibliography{ref}

\end{document}